\documentclass[fleqn,10pt,twocolumn]{ICCAS2020}

\usepackage{cite}
\usepackage{makecell}
\usepackage{graphicx}
\usepackage{amsmath}
\usepackage{float}
\newcommand{\thickhline}{
    \noalign {\ifnum 0=`}\fi \hrule height 1pt
    \futurelet \reserved@a \@xhline
}

\begin{document}

\title{Neural Network-Based Ranging with LTE Channel Impulse Response for Localization in Indoor Environments}

\author{Halim Lee${}^{1}$, Ali A. Abdallah${}^{2}$, Jongmin Park${}^{1}$, Jiwon Seo${}^{1*}$, and Zaher M. Kassas${}^{2, 3}$}

\affils{ ${}^{1}$School of Integrated Technology, Yonsei University,\\
Incheon 21983, Korea (halim.lee, jm97, jiwon.seo@yonsei.ac.kr)\\
${}^{2}$Department of Electrical Engineering and Computer Science, University of California, Irvine, \\
${}^{3}$Department of Mechanical and Aerospace Engineering, University of California, Irvine,\\
California 92697, USA (abdalla2@ucu.edu, zkassas@ieee.org) \\
 ${}^{*}$ Corresponding author}

\abstract{A neural network (NN)-based approach for indoor localization via cellular long-term evolution (LTE) signals is proposed. The approach estimates, from the channel impulse response (CIR), the range between an LTE eNodeB and a receiver. A software-defined radio (SDR) extracts the CIR, which is fed to a long short-term memory model (LSTM) recurrent neural network (RNN) to estimate the range. Experimental results are presented comparing the proposed approach against a baseline RNN without LSTM. The results show a receiver navigating for 100 m in an indoor environment, while receiving signals from one LTE eNodeB. The ranging root-mean squared error (RMSE) and ranging maximum error along the receiver's trajectory were reduced from 13.11 m and 55.68 m, respectively, in the baseline RNN to 9.02 m and 27.40 m, respectively, with the proposed RNN-LSTM.
}

\keywords{long-term evolution (LTE), indoor localization, indoor navigation, recurrent neural network (RNN), long short-term memory (LSTM)
}

\maketitle


\section{Introduction}
Location-based services (LBS) have become an essential part of our lives \cite{GPSWorld14:LBS}. LBS depend on navigation technologies, such as global navigation satellite system (GNSS) \cite{Gogoi19:On,Sun19:Multi-level} and enhanced long-range navigation (eLoran) \cite{Lo10:Improving, Son18:Universal}. LBS can also exploit other radio signals in the environment, such as Wi-Fi \cite{Faragher12:Opportunistic,Yang15:WiFi-based, Khalife15:Indoor} and cellular signals \cite{Gentner12:Particle,Shamaei17:Exploiting,Abdallah18:Machine,delPeral-Rosado18:Survey}.

In outdoor environments, GNSS provide an acceptable localization performance. A receiver's position can be estimated to within a few meters utilizing pseudorange measurements \cite{De10:The,Yoon16:Position}, while decimeter-level accuracy is achievable with carrier phase measurements \cite{Pesyna14:Phase}. However, GNSS are vulnerable to radio frequency interference \cite{Kim17:SFOL, Park18:Dual, Son17:Novel} and atmospheric changes, such as ionosphere anomalies \cite{Seo14:Future, Lee17:Monitoring}. In indoor environments, GNSS signals get severely attenuated, making them practically unusable. For indoor environments, significant attention has been devoted to localization with WiFi \cite{Faragher15:Towards,Zhang20:ASelf}, ultra-wide band (UWB) \cite{Bialer12:Efficient,Xu17:UWB-based}, and radio-frequency identification (RFID) \cite{Yao18:AnIndoor,Zeng19:UHF}.

Cellular signals, particularly long-term evolution (LTE) signals, have shown tremendous promise in circumventing the limitations of GNSS signals in both indoor and outdoor environments \cite{Kassas17:I_Hear}. Cellular signals can be exploited to produce a standalone navigation solution or can be coupled with other sensors (e.g., lidar, inertial measurement unit, etc.). In outdoor environments, recent work has shown that cellular signals could yield meter-level and even lane-level accurate localization on ground vehicles \cite{Yang14:Mobile,Kassas17:LTE,Maaref17:Lane,Kassas17:Robust,Yang20:Mobile} and sub-meter-level accurate localization on aerial vehicles \cite{Khalife18:Precise,Shamaei19:Submeter}. Meter-level accurate localization has been recently reported indoors \cite{Driusso16:Indoor,Gentner18:Channel,Abdallah18:Indoor, Abdallah19:Evaluation,Abdallah20:Deep}. What makes cellular LTE signals especially attractive is their ubiquity, high power, and geometric diversity.

Two types of ranging method are commonly used for LTE-based localization: signal strength-based and time-of-arrival (TOA)-based. Among the two methods, signal-strength-based ranging has the advantage of low complexity. Accordingly, many researchers have studied received signal strength indicator (RSSI)-based ranging methods \cite{Paul09:RSSI, Ahn09:Environmental}.

RSSI is the average of total received power observed in orthogonal frequency-division multiplexing (OFDM) reference symbols \cite{Afroz15:SINR}. Although it is possible to perform ranging using a channel model and RSSI, the accuracy is rather low because of signal reflection and attenuation caused by obstacles. Particularly in indoor environments, because of rapid time-varying channels and the presence of many obstacles, it is difficult to model the channel accurately in real-time, reducing the accuracy of  RSSI-based ranging.

Unlike RSSI, the channel frequency response (CFR) provides detailed information about the channel. LTE  receivers estimate CFR from the cell-specific reference signal (CRS) of the LTE physical layer. The CFR provides information about the channel experienced by each symbol in the frequency domain.

Fingerprinting-based localization using channel information extracted from an LTE downlink signal has been previously studied \cite{Ye17:Neural, Pecoraro18:CSI, Zhang18:Fingerprint-based}. In \cite{Ye17:Neural}, a feed-forward three-layer neural-network-based LTE fingerprinting method was suggested. Eleven channel parameters were used as input to the neural network, which were extracted from the channel impulse response (CIR). The CIR is the inverse Fourier transform of the CFR. In other studies \cite{Pecoraro18:CSI,Zhang18:Fingerprint-based}, channel state information (CSI)-based fingerprinting methods were suggested. CSI is a technical term for the channel response used in IEEE 802.11 a/g/n standards \cite{Yang13:From}. In \cite{Pecoraro18:CSI}, a CSI descriptor-based nearest-neighbor fingerprinting algorithm was presented. The CSI descriptors are composed of elements expressing CSI characteristics, such as the mean, standard deviation, and Fano factors of CSI. In addition, a two-stage cascaded neural network was introduced \cite{Zhang18:Fingerprint-based}.

However, fingerprinting-based localization requires a large number of fingerprinting maps. Considering the difficulty to survey large areas to obtain and maintain the fingerprinting maps, localization based on range measurements from a user to nearby eNodeB is more practical. This paper proposes a recurrent neural network (RNN)-based ranging method using CIR. The CIR was obtained from real LTE signals, and the range between the user equipment (UE) and the LTE eNodeB was estimated. A long short-term memory model (LSTM) network was designed to extract the range from the magnitude of CIR. Experimental results are presented comparing the proposed approach against a baseline recurrent neural network (RNN) without LSTM. The results show a receiver navigating for 109 m in an indoor environment, while receiving signals from one LTE eNodeB. The ranging root-mean squared error (RMSE) and ranging maximum error along the receiver's trajectory were reduced from 13.11 m and 55.68 m, respectively, in the baseline NN to 8.01 m and 31.57 m, respectively, with the proposed RNN-LSTM.

The remainder of this paper is organized as follows. Section 2 describes the proposed RNN-LSTM-based ranging approach. Section 3 presents experimental results in an indoor environment. Section 4 presents concluding remarks.

\section{Proposed Method}\label{Sec:proposed_approach}
This section presents: (i) CIR estimation of the received LTE signals and (ii) proposed RNN-LSTM-based model to estimate ranges from the LTE CIRs.

\subsection{Channel impulse response}\label{subsec:CIR_est}
The LTE system uses orthogonal frequency division multiplexing (OFDM) as a modulation technique with a frame duration of 10 ms and subcarrier spacing $\Delta f = 15$ kHz. In the time-domain, an LTE frame is divided into ten subframes, where each subframe is divided into two slots and each slot consists of seven OFDM symbols. In the-frequency domain, the LTE bandwidth is scalable from 1.4 MHz to 20 MHz with a different number of total and used subcarriers in each configuration.

\begin{figure} [t]
\begin{center}
\includegraphics[width=\columnwidth]{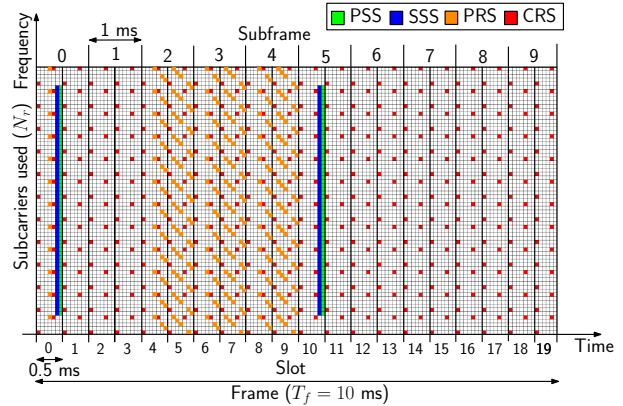}
\caption{LTE frame structure \cite{Abdallah18:Indoor}.} \label{fig:frame}
\end{center}
\end{figure}

CIR estimation can be performed using any of the several LTE reference signals such as: (i) primary synchronization signal (PSS), (ii) secondary synchronization signal (SSS), and (iii) CRS. The PSS and SSS are transmitted to provide the frame start time and the cell ID of the LTE base station, also known as the evolved Node B (eNodeB); however, the PSS and SSS have a fixed bandwidth of 0.93 MHz regardless of the LTE downlink bandwidth. On the other hand, the CRS bandwidth is the same as the transmission bandwidth (i.e., up to 20 MHz). This makes the CRS more attractive for range measurements, especially in multipath environments. The LTE frame structure is shown in Fig. \ref{fig:frame}.

This paper adopts the carrier phase-based software defined receiver (SDR) proposed in \cite{Abdallah19:Performance} in which the CIR is estimated by tracking the CRS. In this receiver, there are two stages: (i) acquisition stage and (ii) tracking stage. In the acquisition stage, the received LTE baseband signal is correlated with all possible locally-generated PSS and SSS sequences to produce a coarse estimate of the frame start time, which is used to control the fast Fourier transform (FFT) window timing. The LTE guard band, also known as cyclic prefix (CP), elements are removed and an FFT is taken to convert the signal into the LTE frame structure. Then, the CIR is estimated using the estimation of signal parameters via rotational invariance techniques (ESPRIT) and used to refine the time-of-arrival (TOA). The phase difference between CFRs estimated from two distinct CRS symbols is used to provide a coarse estimate of Doppler frequency, $\hat{f}_D$.

In the tracking stage, a phase-locked loop (PLL) is implemented to track the phase of the CRS signal. The carrier phase discriminator can be defined as the phase of the integrated CFRs over the entire subcarrier \cite{Shamaei17:LTE}. Then, a second-order loop filter at the output of the discriminator can be used to estimate the rate of change of the carrier phase error, $2\pi \hat{f}_D$, expressed in rad/s. Finally, the TOA estimate, $\hat{e}_{\tau}$, is updated according to
\begin{align} \label{eq:TOA}
  \hat{e}_{\tau}\longleftarrow\hat{e}_{\tau}- \frac{T_{f}}{T_s} v_{\mathrm PLL},
\end{align}
where $T_f = 10$ ms, $T_s$ is the sampling time, and $v_{\mathrm PLL}$  is the output of the PLL.

\subsection{RNN model}
\begin{figure} [t]
\begin{center}
\includegraphics[width=\columnwidth]{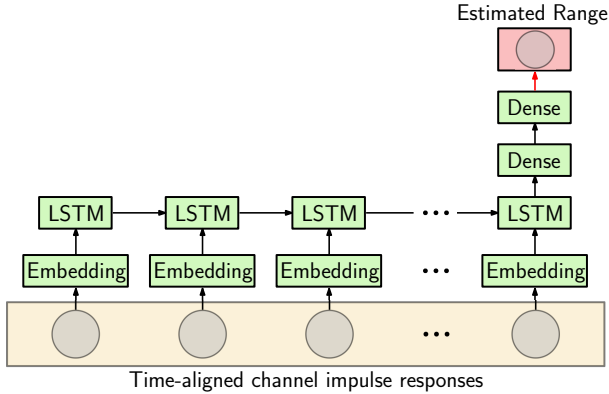}
\caption{Proposed RNN-LSTM structure. The time-aligned CIR and estimated range are the input and output of the proposed RNN-LSTM, respectively. The RNN-LSTM consists of embedding, LSTM, and dense layers.} \label{fig:RNN}
\end{center}
 \vspace{-0.6cm}
\end{figure}

Since the CIR is a temporal sequence, an RNN, which has advantages in solving sequential problems, was designed. The RNN was long short-term memory (LSTM)-based and is depicted in Fig. \ref{fig:RNN}.

The proposed RNN-LSTM consists of one embedding layer, one LSTM layer, and two dense layers. The output dimension of the embedding layer is set to 128. The number of the hidden units of the LSTM layer are set to 128. Further, the output dimensions of the dense layers are set to 128 and 1 for the first and second dense layers, respectively. The rectified linear unit (ReLU) is used as the activation function of dense layers. The output of the last activation function is the estimated range.

\section{Experimental Results}\label{Sec:Exp_results}

\subsection{Experimental Setup and Environmental Layout}
To validate the proposed approach, an experiment was conducted in an indoor environment: Winston Chung Hall building at the University of California, Riverside, USA. Two LTE receivers were used: (i) a rover receiver which navigates indoors and (ii) a base receiver which is placed on the roof of the building. The base receiver has access to GPS and is used to estimate the clock biases of the LTE eNodeB. The estimated eNodeB's clock biases were removed from the rover's measurements; hence, the rover's measurement errors are mainly affected by the multipath. The base receiver was equipped with a single-channel National Instruments (NI) universal software radio peripheral (USRP)-2920 to simultaneously down-mix and synchronously sample LTE signals at 10 Msps. The rover's receiver was equipped with a dual-channel USRP-2954R; however, one channel is exploited here to sample the LTE signals at the sampling rate of 20 Msps. Both receivers were equipped with a consumer-grade cellular omnidirectional antenna to collect LTE data at the carrier frequency of 2145 MHz, which corresponds to the U.S. cellular provider T-Mobile.

\begin{figure} [t]
\begin{center}
\includegraphics[width=\columnwidth]{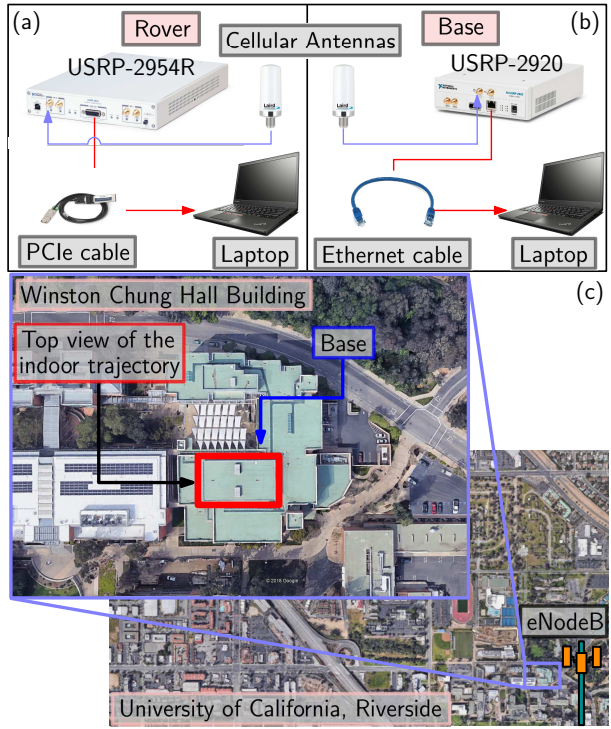}
\caption{Environmental layout and experimental setup. (a) shows the rover's hardware setup, (b) shows the base's hardware setup, and (c) shows the environmental layout, eNodeB location, base location, and rover's trajectory.} \label{fig:Exp_setup_env_layout}
\end{center}
 \vspace{-0.4cm}
\end{figure}

The collected data were processed in a post-processing fashion using the receiver discussed in Subsection \ref{subsec:CIR_est} , where the eNodeB (physical cell ID: 383) was tracked. Throughout the experiment, a smart phone was used to record the location of pre-placed tags at known location on the ground, which was later processed and used as a ground truth. Fig. \ref{fig:Exp_setup_env_layout} shows the environmental layout and the hardware setup for both the rover and the base receivers.

\subsection{RNN Training}
\begin{figure} [t]
\begin{center}
\includegraphics[width=\columnwidth]{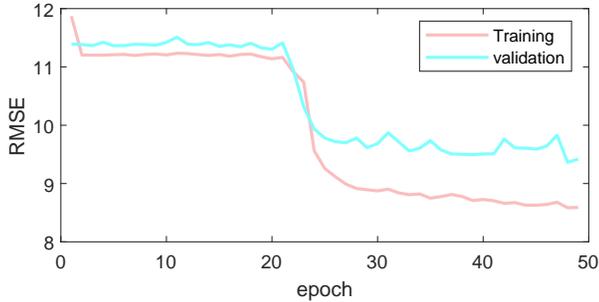}
\caption{Training and validation losses with respect to the training epoch.} \label{fig:Taining_Validation_los}
\end{center}
 \vspace{-0.4cm}
\end{figure}

Throughout the experiment, the receiver travelled a distance of 109 m in 50 seconds. The CRS-based CIR was estimated for each received LTE frame, which provides 5000 samples that were divided as the training, validation, and test data, which were $60\%$, $20\%$, and $20\%$ of the total number of samples, respectively. For the training process, an AdamOptimizer with the learning rate of $10^{-3}$ was used to minimize the mean squared error of the estimated ranges. The LSTM network was implemented on TensorFlow. The learning was conducted using the training and validation data, and the performance was evaluated using the test data. The training and validation losses with respect to the training epoch are shown in Fig. \ref{fig:Taining_Validation_los}. The training and validation losses are defined as the root-mean-squared error (RMSE) of the difference between the estimated range $\hat{r_i}$ and the true range $r_i$, as shown in Eq. (\ref{eq:RMSE}).

\begin{equation} \label{eq:RMSE}
{\mathrm LOSS} = \sqrt{ \frac{1}{n} \sum^n_{i=1} {(\hat{r_i}-r_i)}^2},
\end{equation}
where $n$ is the total number of training or validation samples.

\subsection{Results}
The performance of the proposed RNN-LSTM is compared with that of a baseline RNN. The baseline RNN that we designed for the purpose of comparison consists of two dense layers with the ReLU function as the activation function. The output dimensions of the dense layers were set to 128 and 1 for the first and second dense layers, respectively. The baseline RNN was trained with an AdamOptimizer with the learning rate of $10^{-3}$ for 300 epochs. The only difference between the design of the proposed network and the baseline RNN is the existence of the embedding and LSTM layers for sequential processing. The only difference between the proposed network and the baseline RNN is the existence of the embedding and LSTM layers for sequential processing. The output dimensions, activation function of dense layers, learning rate, and optimizer remain the same for both networks.

\begin{figure} [hbt]
\begin{center}
\includegraphics[width=\columnwidth]{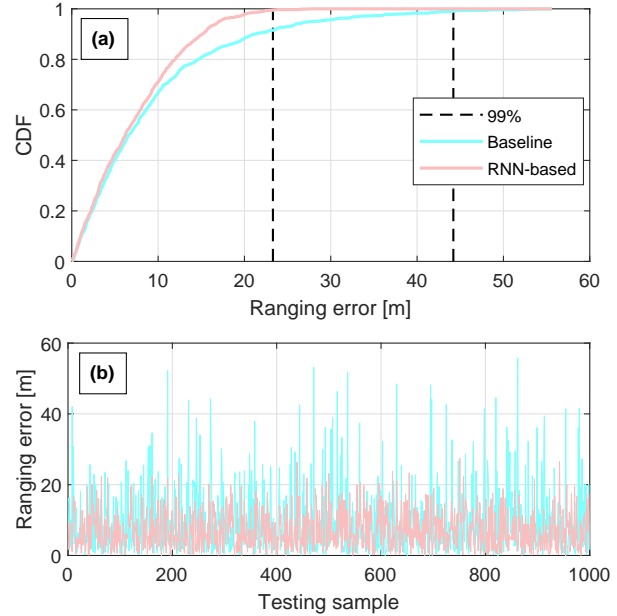}
\caption{(a) CDFs of the ranging errors of the proposed RNN-LSTM and baseline RNN. (b) Ranging errors of both approaches for each test sample out of the 1000 test samples.} \label{fig:CDF_and_Error_plot}
\end{center}
\end{figure}

Fig. \ref{fig:CDF_and_Error_plot} (a) shows the cumulative distribution functions (CDFs) of the ranging errors for the proposed RNN-LSTM and the baseline RNN. Fig. \ref{fig:CDF_and_Error_plot} (b) shows the ranging errors of the proposed RNN-LSTM and the baseline RNN for each test sample. Further, Table \ref{tab:Ranging_results} compares the ranging performance of the two networks. The proposed RNN-LSTM exhibited the ranging RMSE of 9.02 m, outperforming the 13.11 m RMSE of the baseline RNN.

After suspecting that this performance could be caused by overfitting, we tried several approaches that are known to resolve the overfitting problem. First, the learning rate decay \cite{You19:How} was tried to help the optimization process of training. A stochastic gradient descent (SGD) optimizer
with an initial learning rate of 0.1, a decay step of 1000, and a decay rate of 0.96 was used. However, the ranging RMSE of the optimal result obtained by using the learning rate decay was 10.78 m, which is worse than that of the proposed RNN-LSTM without the application of the learning rate decay. Second, we applied the dropout, which is a technique to prevent overfitting by “randomly drop units from the neural network during training \cite{Srivastava14:Dropout}. The ranging RMSE obtained by adding a dropout function between the LSTM layer and the dense layer with the dropout rate of 0.2 was 9.76 m, which is not better than the performance of the proposed RNN-LSTM without the application of the dropout. Finally, we tried more complex layers and then observed overfitting. For example, when the neural network consisted of one embedding layer, one gated recurrent unit (GRU) layer, one LSTM layer, and three dense layers was tested and the learning rate decay was applied, all estimated ranges converged to the same value. The converged value was 332.82 m, which is very close to the mean of the true ranges in the training data (332.67 m). Based on these observations, we concluded that the proposed RNN-LSTM does not suffer from overfitting. Since the data used in this study was collected in a very challenging multipath environment, the data has low generality characteristics.

\vspace{0.1cm}
\begin{table}[h]
\begin{center}
\caption{Ranging Performance Comparison}\label{tab:Ranging_results}
 \vspace{-0.1cm}
\renewcommand{\arraystretch}{1}
 \begin{tabular}{c c c}
 \Xhline{2\arrayrulewidth}
 \multicolumn{1}{c}{\begin{tabular}[c]{@{}c@{}}\textbf{Performance Measure [m]}\end{tabular}} & \multicolumn{1}{c}{\begin{tabular}[c]{@{}c@{}}\textbf{Baseline} \\ \textbf{RNN}\end{tabular}} & \multicolumn{1}{c}{\begin{tabular}[c]{@{}c@{}}\textbf{Proposed}\\ \textbf{RNN-LSTM}\end{tabular}} \\
 \hline
 RMSE & 13.11 & 9.02 \\
 Standard deviation & 9.17 & 5.40 \\
 Maximum error & 55.68 & 27.40 \\
 \Xhline{2\arrayrulewidth}
\end{tabular}
\end{center} 
\end{table}

\section{Conclusion}\label{Sec:Conc}
An RNN-LSTM-based approach to estimate the range between a UE and an LTE eNodeB using the CIR extracted from the CRS signal was developed. An RNN utilizing timely-synchronized magnitudes of CIR was designed. The proposed approach was validated experimentally, where LTE signals were collected in an indoor environment over 109 m in 50 seconds. The proposed approach reduced the ranging RMSE by 68.8\% compared to the performance achieved by a baseline RNN.

\section*{Acknowledgement}
The authors would like to thank Joe Khalife, Kimia Shamaei, and Mahdi Maaref for their help in data collection. This research was supported by the Ministry of Science and ICT (MSIT), Korea, under the High-Potential Individuals Global Training Program (2020-0-01531) supervised by the Institute for Information \& Communications Technology Planning \& Evaluation (IITP) and also supported by the HPC Support Project funded by MSIT and the National IT Industry Promotion Agency (NIPA) of Korea. This work was partially performed under the financial assistance award 70NANB17H192 from U.S. Department of Commerce, National Institute of Standards and Technology (NIST).

\bibliographystyle{IEEEtran}
\bibliography{references}

\end{document}